\documentclass[fleqn,10pt]{wlscirep}
\usepackage[utf8]{inputenc}
\usepackage[T1]{fontenc}
\title{Dynamic Mode Decomposition of \\
       inertial particle caustics in Taylor-Green flow}

\author[1]{Omstavan Samant}
\author[2,*]{Jaya Kumar Alageshan}
\author[3]{Sarveshwar Sharma}
\author[2]{Animesh Kuley}
\affil[1]{Centre for Fusion, Space and Astrophysics, University of Warwick, Coventry, CV4 7AL, UK.}
\affil[2]{Department of Physics, Indian Institute of Science, Bangalore, India - 560012.}
\affil[3]{Institute For Plasma Research, Gandhinagar, Gujarat, India - 382428.}

\affil[*]{jkumar.res@gmail.com}

\begin{abstract}
Inertial particles advected by a background flow can show complex structures. We consider
inertial particles in a 2D Taylor-Green (TG) flow and characterize particle dynamics as a
function of the particle's Stokes number using dynamic mode decomposition (DMD) method from
particle image velocimetry (PIV) like-data. We observe the formation of caustic structures
and analyze them using DMD to (a) determine the Stokes number of the particles, and (b)
estimate the particle Stokes number composition. Our analysis in this idealized flow will
provide useful insight to analyze inertial particles in more complex or turbulent flows.
We propose that the DMD technique can be used to perform similar analysis on an experimental
system.
\end{abstract}
\begin{document}

\flushbottom
\maketitle
%
%
\thispagestyle{empty}

\section{Introduction}

Advection of particles, such as dust or aerosol by a background flow is a ubiquitous
phenomena. And the study of dispersion of these inertial particles are of immense interest both for
applied and natural processes, in particular, to 
analyze oil spills in oceans~\cite{Beron,Mezic,Nencioli,Olascoaga}, dispersion of pollutants
and toxic elements~\cite{Tang, Natusch}, suspended particles in aquatic systems\cite{Espinosa},
formation of clouds~\cite{Shaw,Sapsis} and volcanic plumes~\cite{Haszpra}, and the effect of the
flow patterns generated by the breathing
action and cough on the dispersion of the aerosol particles are crucial to understand the
spread of COVID-19 virus~\cite{COVID1,COVID2,COVID3,COVID4,COVID5,COVID6}.
Numerical studies of inertial particle dispersion in different types of flows, ranging from static~\cite{MR} to turbulent~\cite{Bec,Bec2} flows have shown that the 
particles display complex dynamical
behaviours like formation of fractal clusters\cite{Bec} and caustics\cite{Rama}. The analysis of
the structures formed by the particles encode information about the Stokes number of the particles
and the flow patterns. 

Experimental techniques such as particle image velocimetry (PIV) have been used to track particles
and extract velocity profiles~\cite{PIV} when it is possible to identify individual particles
in an image, but not with certainty to track it between images. If the particle concentration
is so low that it is possible to follow an individual particle it is called particle tracking
velocimetry (PTV). While similar techniques have been adopted to track particles in
simulations, the averaging process reduces the spatial resolution, which is critical
in our application. In simulations, the Osiptov's method~\cite{Healy} have been used to extract
caustic features~\cite{Rama}, which track each particle in PTV like situations but fail for PIV
like data. We propose the use
dynamical mode decomposition (DMD) based scheme to obtain the spatio-temporal particle
distributions as a representative for particle density. DMD methods have been used to extract
coherent structures in simulations and experiments~\cite{DMD1,DMD2,DMD3,DMD4} of fluids.
 We use the DMD method to analyze and extract
the features of the caustics to (a) determine the Stokes number of the particles, and (b)
estimate the relative particle concentrations in a bi-disperse Stokes number system. 
Our approach can also be extended to multiple Stokes number poly-disperse systems.

This paper is organized as follows. In Sec.~\ref{Sec:Model} we present the form of the TG flow,
minimal model of an inertial particle, and the relevant numerical simulation details.
In Sec.~\ref{Sec:Observations} we show the formation of caustic structures and analyze them
using DMD method in Sec.~\ref{Sec:Analysis} and demonstrate how we extract the caustic wavefronts
from the DMD mode. We use the position and the gradient of the wavefront in the DMD eigen mode to
estimate the Stokes number and the composition of a bi-disperse Stokes number systems in
Sec.~\ref{Sec:Results} and present our conclusions in Sec.~\ref{Sec:Conclusions}.

\subsection{Model}
\label{Sec:Model}

We consider a 2D lattice of vortices in the form of a Taylor-Green (TG) flow. 
The TG flow is a steady state solution to the forced, incompressible Navier-Stokes equation
and can be considered a convection model in 2D\cite{Young,Sarracino}. Such a flow can be experimentally
setup using ion solutions in an array of magnets~\cite{Tabeling}. The TG flow is given by the vorticity 
field as
\begin{eqnarray}
   \omega(x,y) = \omega_0 \; \sin\left(\frac{2\pi x}{L}\right) \; 
             \sin\left(\frac{2\pi y}{L}\right)
\end{eqnarray}
and the corresponding velocity field is
\begin{eqnarray}
   \mathbf{u}(x,y) = V_0 
   \begin{bmatrix}
       & \sin\left(\frac{2\pi x}{L}\right) \; \cos\left(\frac{2\pi y}{L}\right) \\
      -& \cos\left(\frac{2\pi x}{L}\right) \; \sin\left(\frac{2\pi y}{L}\right)
   \end{bmatrix}
   \label{eq:vorticity}
\end{eqnarray}
where $x,y \in [0,L)$ and are periodic, $\mathbf{u}$ is
the Eulerian velocity field such that $\omega = \nabla\times\mathbf{u}$.
We choose $V_0$ as the velocity scale and $L$ as the length scale and write
the system parameters in corresponding dimensionless form.
We model the aerosol particles as small rigid spheres, which are effectively points, that 
have density different from the surrounding fluid. The equation of motion of the inertial
particles in a background flow given by the simplified Maxey-Riley
approximation~\cite{MR} for small particles that are much denser than the fluid are
\begin{eqnarray}
   \label{eq:particle}
   \frac{d\mathbf{x}}{dt} &=& \mathbf{v}  \nonumber \\
   \frac{d\mathbf{v}}{dt} &=& \frac{1}{St} \left(\mathbf{u}-\mathbf{v}\right)
\end{eqnarray}
where $St$ is the Stokes number which captures the effect of particle inertia, $\mathbf{x}$ 
is the particle position and $\mathbf{v}$ is the particle velocity (see Appendix~\ref{App1} for
validity of the equations). The case when $St\rightarrow 0$ the particles act as tracers that
follow the velocity stream lines and the eq.~\ref{eq:particle} leads to $\mathbf{v}=\mathbf{u}$.
We use RK4 numerical
scheme to discretize and evolve the particle positions and velocities. Furthermore, we use 
periodic boundary conditions, such that the particles are reintroduced into the system when
they exit the boundary. In our analysis we set the time step to $\Delta t = 0.01
\left(L/V_0\right)$.

\subsection{Observations}
\label{Sec:Observations}

\begin{figure}[!ht]
    \centering
    \includegraphics[scale=0.3]{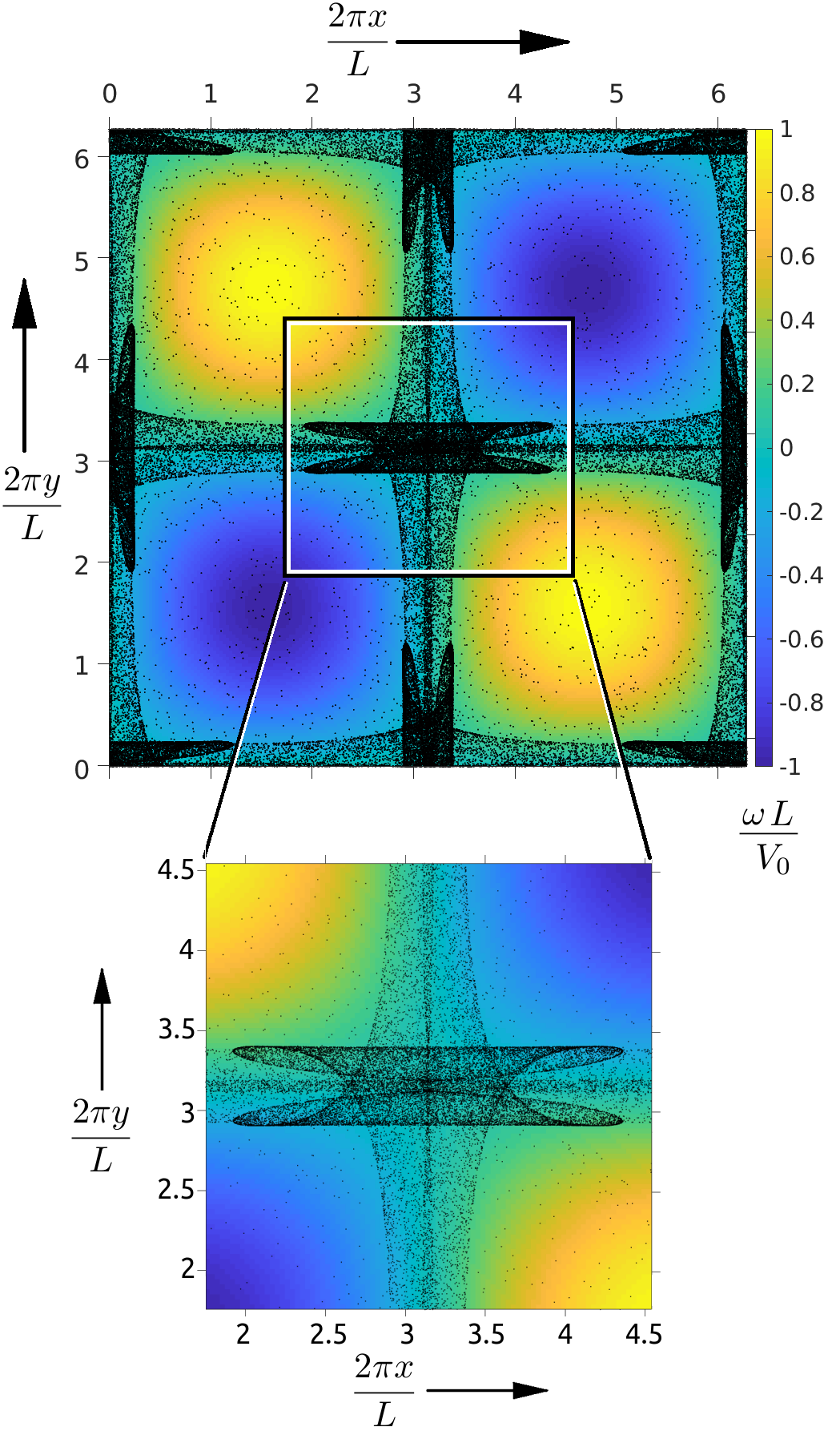}
    \caption{A snapshot of the particle distribution in (a) at $t=750 \: \Delta t$ shows the
       complex structure formed by the mono-dispersed inertial particles with $St=1.0$ in a periodic
       domain of size $L\times L$. The background color-map corresponds to the vorticity
       field and the particles are plotted using black markers. Notice that the particles in the
       high vorticity regions have moved out towards the regions bounding the vortices. The
       zoomed-in version in (b) shows the details of the caustic structures around the central
       region of the domain. See the video~\ref{Mov1} for the evolution of the
       caustic structures.}
    \label{fig:SP_Caustics}
\end{figure}

A typical feature of inertial particles in a background flow is that they are expelled
from high vorticity regions. In a background flow with vortices, the inertial particles can
form caustics\cite{Rama}. So in the case of TG flow, the inertial particles tend to
accumulate along the low vorticity regions that separate the vortices. 
We simulate the dynamics of inertial particles in a TG flow and observe spatial
regions where the Lagrangian velocity field is multi-valued, which are referred to as
caustics~\cite{Caustics1,Caustics2,Caustics3} (see Fig.~\ref{fig:Caustics}(a)).

Figure.~\ref{fig:SP_Caustics} shows the caustic structure formed by a mono-disperse inertial
particle system, when we start the simulation with the particle positions initialized
with uniform random distribution within the $L\times L$ box, and setting the initial velocities
of the particles to zero. 
In the long time limit, the particles accumulate along a curve~\cite{Chaos}.
We observe that the caustics that form in the transient state are robust and stable to
small perturbations (see Appendix~\ref{App1}), whereas the steady state structures break up
and lead to
chaos~\cite{Chaos}. Furthermore, the steady state behaviour strongly depends on the system
size and the boundary conditions.

We find that for a range of Stokes numbers the caustic structures preserve their shapes;
and their sizes depend on $St$. In the following section we use dynamic mode decomposition
(DMD) to extract features of the structures, in particular the caustic wavefront, and study
its size dependence on the Stokes number. Furthermore, the sharpness of the caustic
wavefront enables us to detect and extract their sizes even in presence of poly-disperse
Stokes number systems.

\subsection{Analysis}
\label{Sec:Analysis}

The caustics in Fig.~\ref{fig:SP_Caustics} have a complex structure
and in the presence of multiple Stokes number particles resolving these structures from a
single snap-shot is hard. Therefore we employ the spatio-temporal data in the form of
a video sequence that contains $\mathcal{F}$ frames of $N\times N$ pixel images and analyze
them using the dynamic mode decomposition (DMD) method.

DMD is a data analysis technique that has been used to extract coherent structures in
fluid dynamic systems~\cite{Taira}, where it is able to extract different modes that are
similar to normal modes in linear dynamical systems. The DMD is a data-driven technique introduced 
by Schmid as a numerical procedure for extracting dynamical features from flow data~\cite{Schmid}.
The DMD algorithm takes in a time-series data in the form of vectors $\{\vec{v}_1,\vec{v}_2,...\vec{v}_T\}$
and estimates a linear dynamical system that can generate a map
\begin{eqnarray}
   \vec{v}_{i+1} = \mathcal{A} \; \vec{v}_i
\end{eqnarray}
where $\mathcal{A}$ is an $N^2 \times N^2$ matrix and the eigenvectors of $\mathcal{A}$ form the DMD
modes, with the corresponding eigenvalues.
Finally, the eigenvectors are reshaped into $N\times N$ pixel image to obtain the modes.
A Singular Value Decomposition (SVD) based algorithm for estimating the DMD modes is described
in Appendix~\ref{App2}. 
\begin{figure}[!ht]
    \centering
    \includegraphics[scale=0.22]{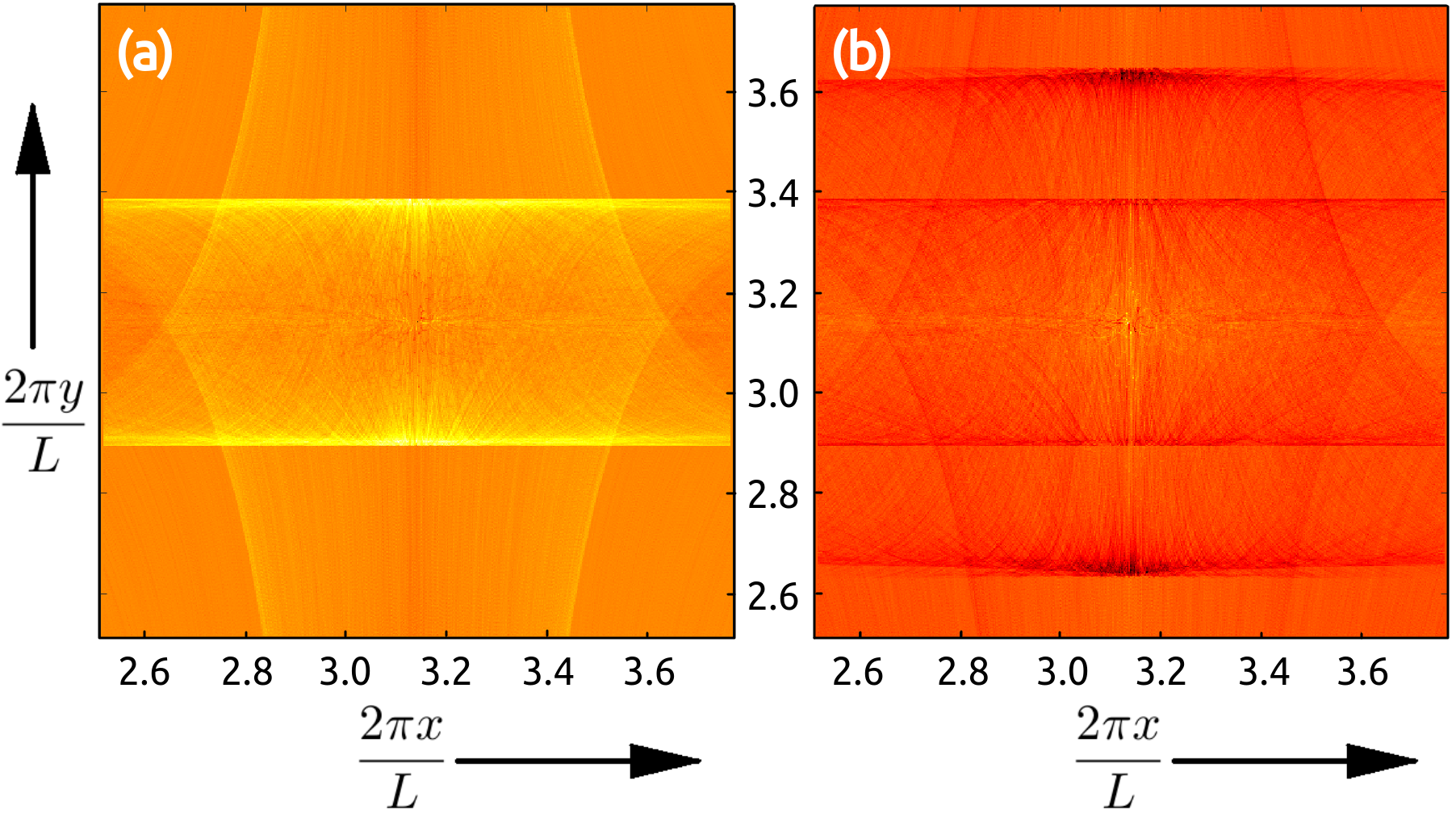}
    \caption{The highest singular DMD eigenvector, $\mathcal{D}^{(1)}$, obtained for:
    (a) $St=1.0$ mono-disperse
    system, (b) a bi-disperse mixture of $St=\{1, 2\}$ with $7:3$ ratio of initial particle concentrations. The horizontal lines correspond to the caustic wavefronts, and the number of
    such fronts indicate the different Stokes number particles. Also notice that the wavefronts
    corresponding to $St=1.0$ are vertically aligned in both (a) and (b), indicating that the
    positions of DMD wavefronts are not perturbed by the presence of particles with different
    Stokes numbers.}
    \label{fig:SP_DMD}
\end{figure}

\begin{figure}[!ht]
    \centering
    \includegraphics[scale=0.35]{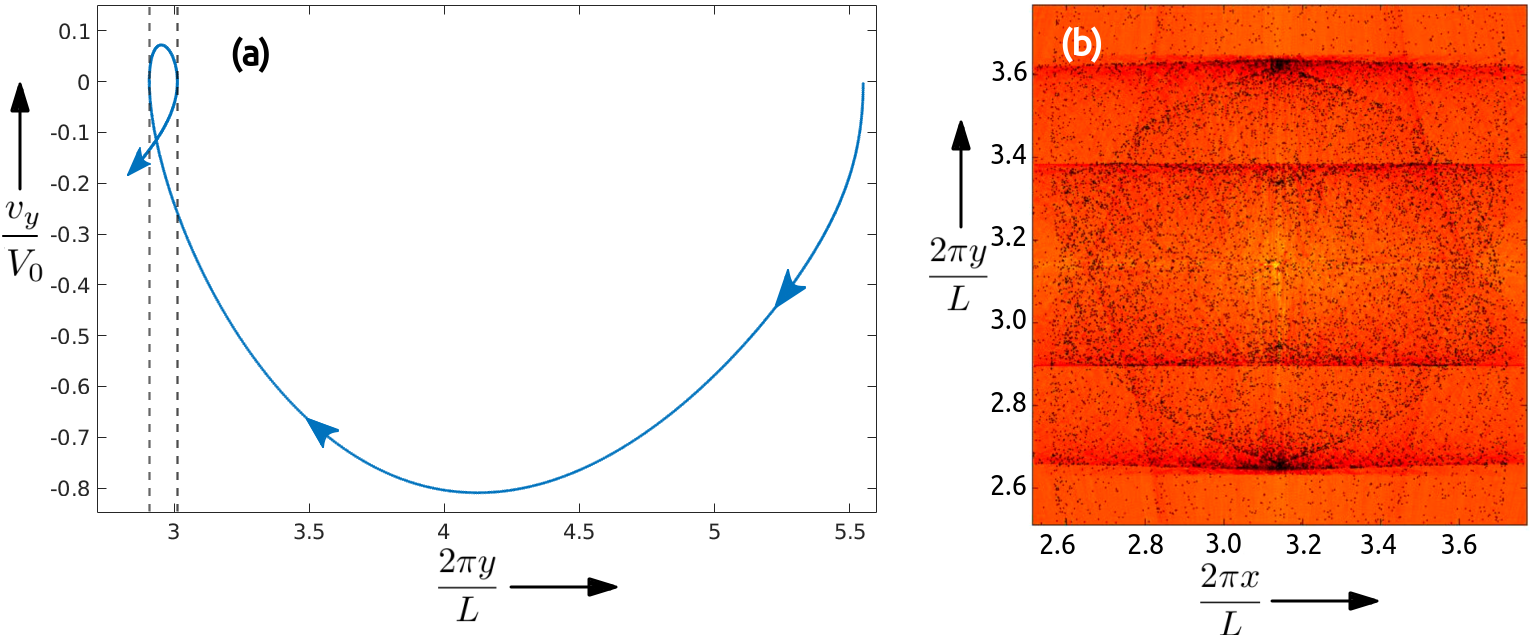}
    \caption{(a) The particle trajectory in the reduced phase space, $(y,v_y)$, shows the
         multi-valued nature of the casutics in velocity between the dotted vertical 
         lines~\cite{Caustics3}. The movie in~\ref{Mov2} shows the
         evolution of the bi-disperse particles overlaid on the corresponding DMD from
         Fig.~\ref{fig:SP_DMD}(b) and the plot (b) shows a snap-shot at the $650^{th}$ iteration
         step when the caustics and the DMD wavefronts coincide. Notice that the first DMD
         picks out only the slow moving horizontal caustics.}
    \label{fig:Caustics}
\end{figure}

Let $i$ stand for the iteration number such that the particles are in their stationary initial
state and start their evolution at $i=0$. Then for our data, the vectors $v_i$ are obtained by
rearranging the $N\times N$ pixel images at instant $i$ into $N^2\times 1$ vector. 
In our DMD analysis we employ $\{v_{250}, v_{251},...v_{750}\}$ (i.e. $\mathcal{F}=500$), as the
modes obtained 
are the sharpest for this range. Let $\mathcal{D}^{(\alpha)}(j,k)$ represent the $(j,k)^{th}$
pixel of the $\alpha^{th}$ eigenmode, ordered in terms of decreasing absolute eigenvalue.
Since the caustics are localized around the central region of the domain, we use a zoomed-in
region of size $512 \times 512$ pixels (i.e. $N=512$) in our analysis, as shown in
Fig.~\ref{fig:SP_Caustics}. We find that the highest singular
eigenvalue mode, namely $\mathcal{D}^{(1)}$, shown in Fig.~\ref{fig:SP_DMD}(a) highlights
a straight-line caustic structure, which we refer to as the wavefront. The eigenvalues of other
DMD modes decay exponentially. We observe that the position of the wavefront in $\mathcal{D}^{(1)}$
has a systematic dependence on the Stokes number, and to extract this relation we detect the
location of the wavefronts using edge detection techniques. Similarly, when we perform DMD
analysis on a bi-disperse system $\mathcal{D}^{(1)}(j,k)$ shows two distinct wavefronts
(see Fig.~\ref{fig:SP_DMD}(b)) corresponding to the two different Stokes numbers; and
here DMD uses the velocity information to unambiguously extract the wave fronts. 
In particular, Fig.~\ref{fig:Caustics}(a) shows the reduced phase space portrait
of a typical particle which form the caustic~\cite{Caustics3} and Fig.~\ref{fig:Caustics}(b)
shows the particles overlaid on top of the DMD that demonstrates the DMD's ability 
to extract the caustic structures.
Furthermore, we find that the intensity
of each wavefront compared to the background, which we refer to as prominence~\cite{Pro}, depends
on the corresponding initial particle concentrations in the system.

We now prescribe a method to extract the position of the wavefront from DMD. We use a
Sobel operator~\cite{Sobel} such that the vertical gradient of the first DMD mode is
given by
\begin{eqnarray}
   G(j,k) \; = \; \frac{\partial \mathcal{D}^{(1)}(j,k)}{\partial y} 
          \; \approx \; \frac{1}{\Delta y} \begin{bmatrix}  
                                    \;\;1 &  \;\;2 &  \;\;1 \\
                                    \;\;0 &  \;\;0 &  \;\;0 \\
                                   -1 & -2 & -1  \end{bmatrix}
                    \circledast \; \mathcal{D}^{(1)}(j,k)
\end{eqnarray}
where $\circledast$ represents the 2D convolution operator~\cite{Conv}, and $\Delta y$ is the spacing
in DMD along the $y$-axis. We then sum over the values in the
x-direction to get a 1D function of y as
\begin{eqnarray}
   \langle G \rangle_x = \int_0^L G(j,k)\; dx \;\; \approx \; \sum_{j=1}^{N} G(j,k) \; \Delta x
\end{eqnarray}
where $\Delta x$ is the spacing in DMD along the $x$-axis, and we choose a square grid with 
$\Delta x = \Delta y$ such that $\langle G \rangle_x$ is by definition independent of the grid spacing
and is dimensionless.
In the next section we describe how the location and the value of the peaks in $\langle G \rangle_x$
can be used to find the Stokes number of the particles and the relative initial concentrations in
the case of a bi-disperse system.

\section{Results}
\label{Sec:Results}

\subsection{Determination of Stokes number from DMD}
\begin{figure}[!ht]
    \centering
    \includegraphics[scale=0.4]{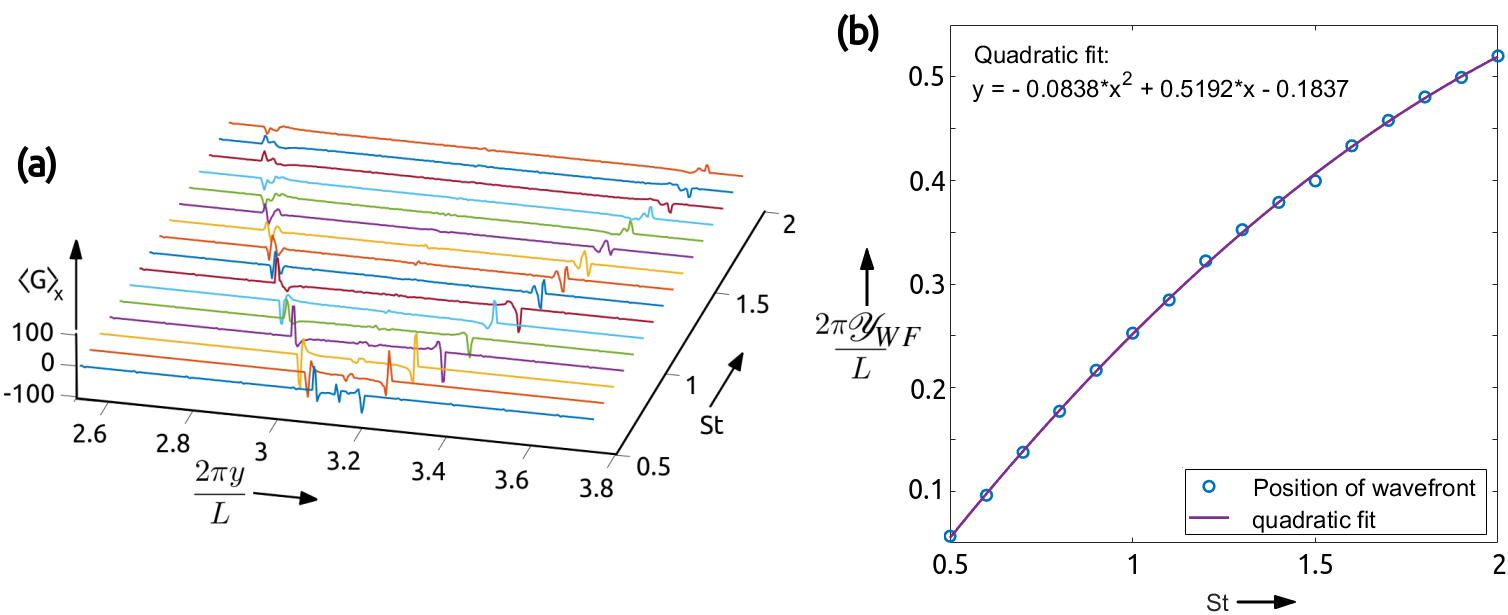}
    \caption{In (a) we show the plots of $\langle G \rangle_x$ obtained for different values
    of the Stokes number from mono-disperse systems. We extract the location of the peaks in
    $\langle G \rangle_x$ for each $St$, as defined in eq.~\ref{eqn:WF}, to generate the plot
    in (b) and we find that the $\mathcal{Y}_{WF}$ and $St$ have a quadratic dependence.}
    \label{fig:St_St}
\end{figure}
The $\langle G \rangle_x$ is obtained from the DMD as described in Sec.~\ref{Sec:Analysis} by
simulating mono-disperse Stokes number particle systems to generate the plots in Fig.~\ref{fig:St_St}(a)
which shows
$\langle G \rangle_x$ as a function of $St$. The alignment of the peaks in $\langle G \rangle_x$ along a
curve indicates a systematic dependence of the location of wavefront on the Stokes number. To extract
this relation we first get the location of the caustic wavefront from the 
domain center using the position of the peaks in $\langle G \rangle_x$ given by
\begin{eqnarray}
   \centering
   \mathcal{Y}_{WF} = \arg\max_y\langle G \rangle_x - \frac{L}{2}
   \label{eqn:WF}
\end{eqnarray}
where $\arg\max_y$ gives the value of $y$ for which $\langle G \rangle_x$ is maximum. We
then plot $\mathcal{Y}_{WF}$ as a function of $St$ as shown in Fig.~\ref{fig:St_St}(b). Using a
non-linear least squares fit method we find that the relation is of the form $\mathcal{Y}_{WF}
\sim a \; St^2 + b \; St + c$, with values of the parameters $a$, $b$, and $c$ as indicated in
Fig.~\ref{fig:St_St}(b), where $x$ represents $St$. Now,
extrapolating the fit we find that $\mathcal{Y}_{WF}=0$ at $St=0.3767$ and becomes multi-valued
for $St>3.0979$, thus setting the limits on the validity of the relation. The three parameters
in the relation can be estimated experimentally using calibrated measurements and the relation can
be used to predict the $St$ of new particle systems. In particular we demonstrate that the above
method can be generalized to work in case of bi-disperse system.

\begin{figure}[!ht]
    \centering
    \includegraphics[scale=0.35]{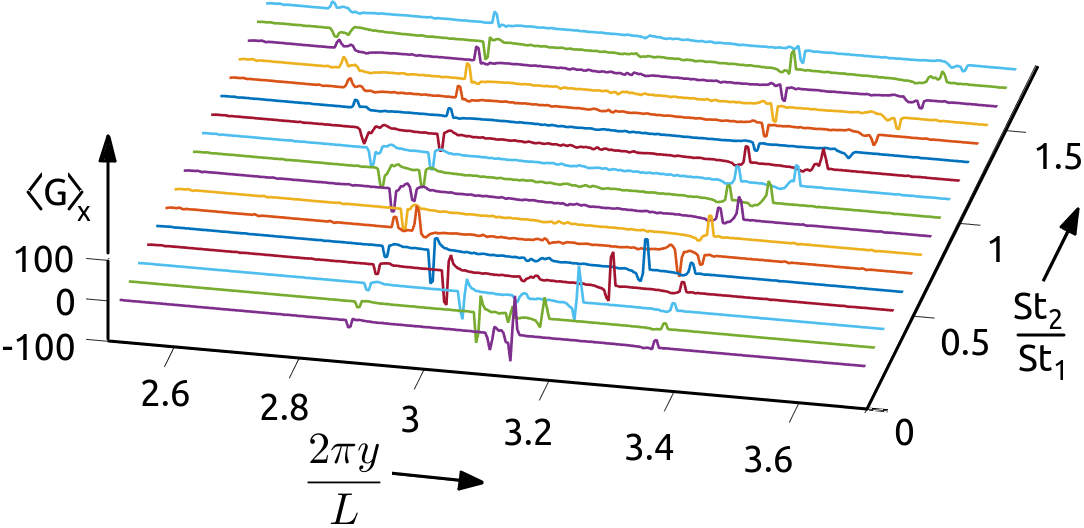}
    \caption{Plots of $\langle G \rangle_x$ for a bi-disperse system, with one of the Stokes number
    fixed at one ($St_1=1.0$), and varying $St_2$. Notice that the peaks in $\langle G \rangle_x$
    corresponding to $St_1$ are aligned at the same location along $y$, whereas the peaks
    due to $St_2$ show similar trend as the plots for mono-disperse systems in Fig.~\ref{fig:St_St}(a).}
    \label{fig:edge_loc_plot_fit}
\end{figure}

As shown in Fig.~\ref{fig:SP_DMD}(b), for a bi-disperse system the DMD has two sets of caustic
wavefronts, corresponding to each Stokes number. Now we set one of the Stokes number fixed at one 
($St_1 = 1.0$), vary the $St_2$ and find $\langle G \rangle_x$ to generate the plots in 
Fig.~\ref{fig:edge_loc_plot_fit}. The results in Fig.~\ref{fig:edge_loc_plot_fit} show that even in the
bi-disperse system the caustic wavefront has the same characteristic behaviour on the Stokes number as
the mono-disperse system. In particular, the wavefront corresponding to $St_1$ has a fixed location and
the wavefront due to $St_2$ preserves the dependence on $\mathcal{Y}_{WF}$ of the mono-disperse system.
Our studies with poly-disperse $St$ systems show that the caustic wavefronts can be used to find the
Stokes number of different particles in the system using the relation obtained from a mono-disperse
system. 

\subsection{Estimation of particle concentration}
\begin{figure}[!ht]
    \centering
    \includegraphics[scale=0.35]{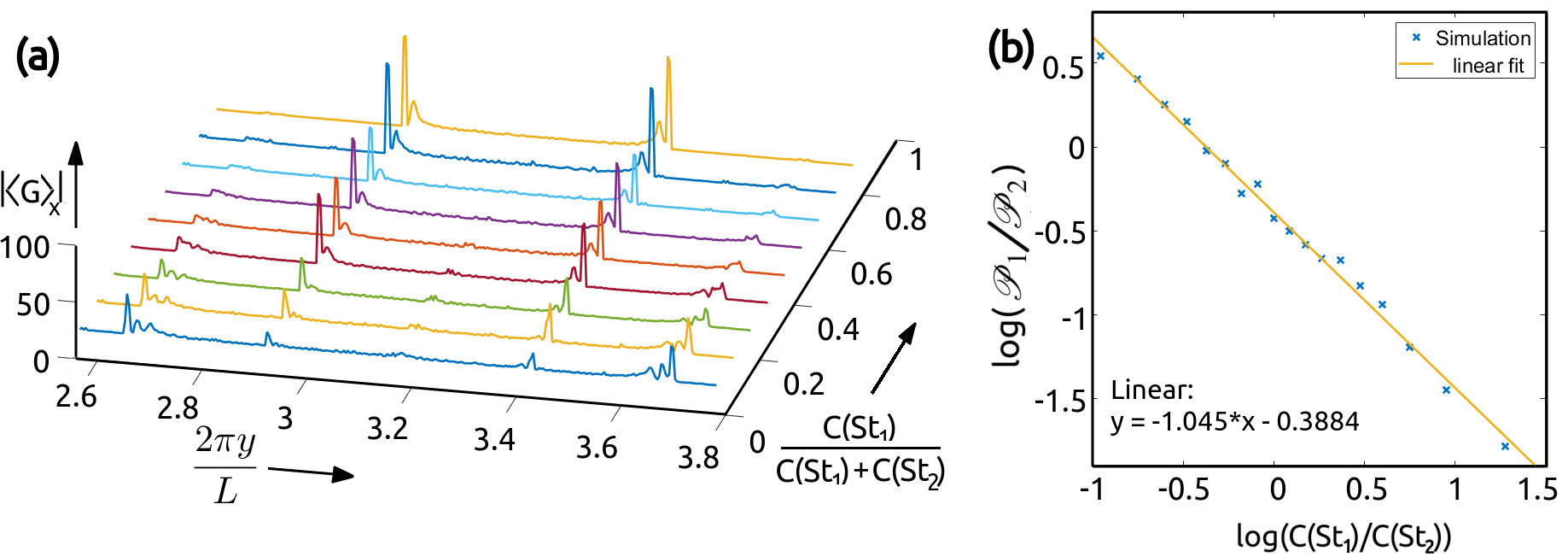}
    \caption{The plot (a) shows the variation in the absolute value of $\langle G \rangle_x$ for a 
    bi-disperse system, with $St_1=1$ and $St_2=2$ particles, for various initial concentration
    fraction $C(St_1)/(C(St_1)+C(St_2))$. Notice that the peaks corresponding to each wavefront is not
    unique and have a finite spread in y. In (b) the relation between the prominence corresponding to
    $St_1$ and $St_2$ are given by $\mathcal{P}_1$, $\mathcal{P}_2$ respectively, as a function of the
    initial particle concentrations $C$ is shown in a log-log plot. The linear fit shows that the ratios
    of the peaks of $\hat{|\langle G \rangle|}$ and the ratios of the concentration are related by a
    power-law, with a power close to -1.}
    \label{fig:St_Composition}
\end{figure}

Until now the bi-disperse systems that we considered had equal number of $St_1$ and $St_2$ particles,
with uniform initial distribution in space. Now we study the variation in $\mathcal{D}^{(1)}$ w.r.t. the
change in relative number of particles. We observe that the intensity of the wavefront or the gradient
in the DMD image depends on the number of particles or the initial uniform concentration, denoted by
$C(St)$. 

The variable $\langle G \rangle_x$ gives the gradient of $\mathcal{D}^{(1)}$ along the vertical direction
and the magnitude of
the gradient indicates sharpness of the wavefronts (see Fig.~\ref{fig:St_Composition}(a)). To measure the
sharpness of the wavefront we define a "prominence" parameter, $\mathcal{P}$, as the sum of the non zero
values of $\Bar{|\langle G \rangle_x|}$ in the neighbourhood of the wavefront, which takes into account
multiple peaks in the vicinity of the wavefront. We find that the prominence of the wavefront 
has a systematic dependence on the initial concentration of the corresponding Stokes number particles
and from Fig.~\ref{fig:St_Composition}(b) we find that on a log-log plot the relation is linear, with
a slope approximately equal to -1. This implies that in a bi-disperse system the ratio of the prominence
is inversely related to the ratio of initial concentrations. We can use this relation to predict the
concentration of various Stokes number particles in a system.

\section{Conclusions}
\label{Sec:Conclusions}

We study the dynamics of inertial particles in a Taylor-Green flow with periodic boundary
conditions in 2D. In a minimal model of inertial particles we observe that for a mono-disperse
Stokes number system, starting from a
uniform distribution of stationary particles, the particle distribution forms caustics 
in the strain dominated region of the flow. 
We use the DMD method to analyze the PIV-like
time-series data of the spatio-temporal particle distribution and find that the largest
absolute eigenvalue mode is effective in extracting the caustic wavefront-like structure.
We notice that
(a) the position of the wavefront depends on the particle Stokes number and employ standard
image processing techniques to quantitatively extract a quadratic relation. Using this
relation we can predict the Stokes number from the wavefront position. Furthermore, we 
find that for a bi-disperse system the DMD is able to extract two different wavefronts 
corresponding to each Stokes number and the positions of each wavefront follow the same
quadratic relation as in the case of mono-disperse system. We also observe that (b) the
sharpness of the wavefront in the DMD, measured in terms of prominence, depends on the initial
particle concentration and find that for a bi-disperse Stokes number system the ratio of
the wavefront prominence is inversely proportional to the corresponding Stokes number
initial concentration. Hence the measurement of prominence can be used to estimate the
concentration of the corresponding Stokes number particles. 

We propose that the DMD technique can be used to analyze real experimental PIV data of caustics
and perform similar analysis to extract information about the Stokes numbers and concentrations
of the particles. In future we will consider detailed Navier-Stokes equation for the
self-consistent evolution of the velocities and analyze the caustic structures in turbulent flows.

\appendix

\section{Validity of particle dynamics}
\label{App1}

The particle dynamics we use in our study is a special case of~\cite{MR}
\begin{eqnarray}
   \frac{d\mathbf{x}}{dt} &=& \mathbf{v}  \nonumber \\
   \frac{d\mathbf{v}}{dt} &=& \frac{1}{St} \left(\mathbf{u}-\mathbf{v}\right)
      + R \; \left( \mathbf{u}.\nabla\mathbf{u} + \frac{1}{2} \mathbf{v}.\nabla\mathbf{u} \right)
      + \mathbf{\eta}(\mathbf{v},t)
   \label{eq:particle2}
\end{eqnarray}
where $\eta$ is the white noise and $R=\rho_f/(\rho_p+\rho_f/2)$ for density of particle $\rho_p$
and density of fluid $\rho_f$.
We present our results for $\eta=0$ and $R=0$,
and use small values of these parameters to verify the stability of our results.

\section{DMD algorithm}
\label{App2}

The $N\times N$ pixel image at $k^{th}$ instant, $I^k(i,j)$, is rearranged into an $N^2\times 1$
vector $X^k(m)$ (note that we subtract the mean value from $X^k(m)$). Now the $\mathcal{F}$ sequence
of image frames are appended together to form $N^2 \times
N_T$ matrix $Y$. Let the $N^2 \times (\mathcal{F}-1)$ dimensional matrix formed from the first 
$(\mathcal{F}-1)$ frames be $Y_1$ and the last $(\mathcal{F}-1)$ frames be $Y_2$, i.e.
\begin{eqnarray}
   \left[ I^k \right]_{N\times N} & \longrightarrow & \left[X^k\right]_{N^2 \times 1} \nonumber \\
   \left[ X^1  | X^2  | ... | X^{N_T} \right] & \longrightarrow & \left[ Y \right]_{N^2 \times N_T} 
                                         \nonumber \\
   \left[ X^1  | X^2  | ... | X^{(N_T-1)} \right] & \longrightarrow & \left[ Y_1 
                                      \right]_{N^2 \times (N_T-1)} \nonumber \\
   \left[ X^2  | X^3  | ... | X^{N_T} \right] & \longrightarrow & \left[ Y_2 \right]_{N^2 \times (N_T-1)}
\end{eqnarray}

We find the singular value decomposition (SVD) of the matrix $Y_1$,
such that $Y_1 = U\Sigma V^*$, where $U$ is a $N^2\times N^2$ complex
unitary matrix, $\Sigma$ is an $N^2 \times (\mathcal{F}-1)$ rectangular diagonal matrix with
non-negative real numbers on the diagonal, and $V$ is a $(\mathcal{F}-1)\times(\mathcal{F}-1)$ real
or complex unitary matrix.

Now choose a lower dimensional SVD matrices made up of first $n_T (<<\mathcal{F})$ columns, represented
by $\Tilde{U}$, $\Tilde{V}$, and the first $n_T\times n_T$ block of $\Sigma$ as $\Tilde{\Sigma}$.
Define a matrix $n_T\times n_T$ matrix $A$ as
\begin{eqnarray}
   A = \Tilde{U}^* \; Y_2 \; \Tilde{V} \; \Tilde{\Sigma}^{-1}
\end{eqnarray}

Then the dynamic modes are the $N^2 \times 1$ eigenvectors of matrix $A$ that is rearranged into
$N\times N$ matrix.

\section{Movies}

The movies referred to in the text are available in the additional materials.

\subsection{See "Vid1.mp4"}
\label{Mov1}

\subsection{See "Vid2.mp4"}
\label{Mov2}

\pagebreak

\nocite{*}

\section*{Acknowledgements}

This work is supported by Board of Research in Nuclear Sciences (BRNS Sanctioned no.
39/14/05/2018-BRNS), Science and Engineering Research Board EMEQ program (SERB Sanctioned
no. EEQ/2017/000164), Infosys Foundation Young Investigator Award, and IISc startup grant.
We thank Rahul Pandit for valuable inputs. JKA acknowledges useful discussions with Akhilesh
Kumar Verma.

\section*{Author contributions statement}
J.K.A, S.S., and A.K. conceived the problem. O.S. and J.K.A. performed simulations. All authors analysed the results. O.S. and J.K.A. wrote the main manuscript text. All authors reviewed the manuscript.

\section*{Additional information}

\subsection*{Corresponding author}
Correspondence to Jaya Kumar Alageshan (jkumar.res@gmail.com).

\subsection*{Competing interests}
The authors declare no competing interests.

\end{document}